\documentclass[manuscript]{aastex}

\usepackage[dvips]{color}

\slugcomment{}

\shorttitle{short title}
\shortauthors{Matsumoto et al.}

\begin{document}



\title{The Orbital Stability of Planets Trapped in the First-Order Mean-Motion Resonances}


\author{Yuji Matsumoto, \altaffilmark{1}\altaffilmark{, 2}}
\email{ymatsumoto@geo.titech.ac.jp}

\author{Makiko Nagasawa\altaffilmark{1}}
\email{nagasawa.m.ad@m.titech.ac.jp}

\and

\author{Shigeru Ida\altaffilmark{1}}
\email{ida@geo.titech.ac.jp}


\altaffiltext{1}{Department of Earth and Planetary Sciences, Tokyo Institute of Technology, Ookayama, Meguro-ku, Tokyo 152-8551, Japan}
\altaffiltext{2}{+81-3-5734-2340}


\begin{abstract}
Many extrasolar planetary systems containing multiple super-Earths have been discovered. 
N-body simulations taking into account standard type-I planetary migration suggest that protoplanets are captured into mean-motion resonant orbits near the inner disk edge at which the migration is halted. 
Previous N-body simulations suggested that orbital stability of the resonant systems depends on number of the captured planets. 
In the unstable case, through close scattering and merging between planets, non-resonant multiple systems are finally formed. 
In this paper, we investigate the critical number of the resonantly trapped planets beyond which orbital instability occurs after disk gas depletion. 
We find that when the total number of planets ($N$) is larger than the critical number ($N_{\rm crit}$), crossing time that is a timescale of initiation of the orbital instability is similar to non-resonant cases, while the orbital instability never occurs within the orbital calculation time ($10^8$ Kepler time) for $N\leq N_{\rm crit}$. 
Thus, the transition of crossing time across the critical number is drastic. 
When all the planets are trapped in 7:6 resonance of adjacent pairs, $N_{\rm crit} = 4$. 
We examine the dependence of the critical number of 4:3, 6:5 and 8:7 resonance by changing the orbital separation in mutual Hill radii and planetary mass. 
The critical number increases with increasing the orbital separation in mutual Hill radii with fixed planetary mass and increases with increasing planetary mass with fixed the orbital separation in mutual Hill radii. 
We also calculate the case of a system which is not composed of the same resonance. 
The sharp transition of the stability can be responsible for the diversity of multiple super-Earths (non-resonant or resonant), that is being revealed by $Kepler$ mission.

\end{abstract}





\section{Introduction}

Seventeen years searchers of extrasolar planets have found more than 100
super-Earths and hot Neptunes ($\leq 30M_{\oplus}$). 
Some of them form multiple planet systems. 
$Kepler$ Mission reports 885 of those multiples in 361 systems (Batalha $et\ al$. 2012; Borucki $et\ al$. 2011a, 2011b). 
Period ratios of pairs of planets show some peaks at commensurable ratios (Lissauer $et\ al$. 2011; Fabrycky $et\ al$. 2012). 
Since orbital angles such as an argument of pericenter are unknown in many of the systems, 
it is not clear that the planets are actually in the mean-motion resonances. 
In data of first four months of $Kepler$ Mission, 
Among 158 multi-super Earth candidates and multi-Neptune candidates, about 25\% planets have orbital periods which are almost commensurable with neighboring planets within 3\% period ratio. 
Veras and Ford (2012) identified 70 non-resonant KOI ($Kepler$ Objects of Interest) near resonant pairs which are not in a mean-motion resonance. 
Terquem and Papaloizou (2007) proposed a mechanism to form resonant planets near the disk inner edge. 
Possible mechanisms for small derivations from the resonance were also raised by Papaloizou and Terquem (2010).

When planets are in the mean-motion resonance, the conjunction periods of the planets are expressed as the integer ratio of periods of each planet. 
That means, conjunctions of the planets always occur the same relative positions. 
Planets in a resonance can become stable depending on the configuration of conjunctions. 
For example, Neptune-Pluto are in 3:2 mean-motion resonance. 
Although the orbits of them cross, they always avoid close approach. 
Neptune-Pluto system are long-term stable (Cohen and Hubbard 1965).

In a gas disk, growing protoplanets migrate toward their 
central stars due to type-I migration. 
The migration is stopped when a protoplanet arrives at the inner edge of the gas disk, often assumed to be at the corotation point of the star ($\sim 0.05-0.1$ AU), if it exists. 
N-body simulations (Terquem and Papaloizou 2007; Ogihara and Ida 2009) showed that in the case of standard type-I migration (e.g., Tanaka $et\ al$. 2002) 
subsequently migrating protoplanets are usually not trapped in a mean-motion resonance at a first encounter with the protoplanet at the disk edge. 
After some close scatterings and collisions, they are eventually trapped in resonances because they are subject to relatively slow type-I migrations near the edge and eccentricity damping after relaxation. 
Through merging of many planets that have migrated to the inner edge, only several merged bodies finally remain in mean-motion resonant orbits. 
Although several inner planets are pushed inside of gas disk edge ($\lesssim0.05$ AU) and others are in the gas disk ($\gtrsim0.05$ AU), they keep the relation of mean-motion resonances. 
These planets are spaced by $5-9$ Hill radius with each other and stay stable even after gas depletion in which eccentricity damping no more operates. 

However, Ogihara and Ida (2009) have found that in the case of migration 
that is slow compared to the rates predicted for standard type-I migration, orbital evolution is totally different and final orbital configuration is non-resonant. 
In this case, subsequently migrating protoplanets are trapped in mean-motion resonances. 
As a result, about $40$ small protoplanets queue in low order mean-motion resonances having closer separations in the gas disk at $\gtrsim0.05$ AU. 
Few of inner planets are pushed into the inner cavity, because of "eccentricity trapping" caused by torque balance between migration torque and edge torque (Ogihara $et\ al$., 2010). 
In contrast to the fast migration case, the planets become orbitally unstable after the gas depletion, i.e., their eccentricities are excited and their orbits start crossing. 
Through collisions and merges of planets, they are kicked out of the resonances and finally several planets are formed in non-resonant orbits with the large orbital separation ($\sim15-20$ Hill radius). 
Although resonant systems are generally more stable than non-resonant systems, results of Ogihara and Ida (2009) showed that multi-planet systems whose planets are initially in overpopulated resonant orbits become unstable in relatively short timescale after gas depletion and end up with dynamically relaxed non-resonant systems. 

Although the crossing time at which multi-planet systems have been extensively studied in gas free environment by N-body simulations (e.g., Chambers $et\ al$., 1996; Yoshinaga $et\ al$., 1999; Zhou $et\ al$., 2007), they only investigated non-resonant systems. 
Because of type-I migration and eccentricity damping, the systems that we consider are deep in resonances, so that the crossing time can be very different from that found by the previous studies.

Although the previous studies on crossing times were concentrated on non-resonant systems, 
the observed resonant multi-super Earths' systems near stars and N-body simulations suggest the occurrence of resonance trapping as a consequence of planetary migrations. 
In this paper, we mainly consider high-integer resonance e.g., 6:5 or 7:6. 
This is because previous N-body simulations suggest that 
proto-super-Earths are once in a close resonances with separations of $\sim 5 - 9$ Hill radius 
and cause instability due to overpopulation later on. 
We calculate the crossing time of systems in first-order mean-motion resonances, 
by changing the total number of plants, the orbital separation in mutual Hill radii, and planetary masses. 
In \S\ref{previous}, we summarize previous studies of N-body simulations to evaluate the crossing time, $t_{\rm cross}$. 
In \S\ref{num_model}, we explain numerical models of our simulations. 
As commensurability of orbital periods does not necessarily mean mean-motion resonance, 
in \S\ref{results}, we study both cases that resonant and non-resonant systems having the same orbital commensurability. 
We discuss the results in \S\ref{conclusions}.

\section{Previous Studies on Crossing Time of Multi-Planet Systems} \label{previous}
Here we summarize previous studies of N-body simulations to evaluate $t_{\rm cross}$ for non-resonant systems in order to make clear the purpose of our simulations. 
%
%
%
%
Chambers $et\ al$. (1996) first investigated the crossing time of multi-planet systems, at which the first close encounter occurs. 
They performed orbital calculations of equal mass protoplanets with various mass from $10^{-9}M_{\odot}$ to $10^{-5}M_{\odot}$. 
They put planets on initially circular and coplanar orbits with mutual separations $a_j - a_i = Kr_{{\rm H}i,j}$, where $r_{{\rm H}i,j}$ is the mutual Hill radius of planets $i$ and $j$, setting the innermost planet at $a_1 = 1$ AU.
They repeated orbital simulations 3 times for the same Hill separation $K$ changing planetary longitudes. 
The calculations were continued until the first encounter, which is defined by the time when distance between two planets becomes smaller than one mutual Hill radius occurs. 
They found that $t_{\rm cross}$ is given approximately by an empirical relation, 
\begin{eqnarray}
 \log{t_{\rm cross} } = b K + c,
 \label{eq:Chambers+}
\end{eqnarray}
where $b$ and $c$ are constants. 
When systems are composed of 3 planets whose mass are $10^{-7}M_{\odot}$, $b=1.176$ and $c=1.663$ for example. 
The values of these constants depend on planetary mass ($M_{\rm p}$) and the number of planets ($N$). 
But when a system has more than 5 planets, 
adding further planets does not make significant difference to the stability of the system. 

%
The crossing time of protoplanets also depends on initial eccentricities and inclinations of protoplanets (Yoshinaga $et\ al$. 1999; Zhou $et\ al$. 2007).
Yoshinaga $et\ al$. (1999) found that the two constants $b$ and $c$ decrease proportional to root mean square of eccentricities and inclinations. 
The dependence of constants $b$ and $c$ on planetary mass, $M_{\rm p}$, was studied by Duncan and Lissauer (1997) and Zhou $et\ al$. (2007). 
Duncan and Lissauer (1997) studied the crossing time of Uranian satellite system with multiplied satellite mass and Zhou $et\ al$. (2007) investigated the crossing time of protoplanetary systems whose settings are similar to Chambers $et\ al$. (1996). 
From numerical calculations of the crossing time with different masses, they concluded $\log{t_{\rm cross}}\propto \log{M_{\rm p}}$. 
They also empirically expressed dependence on eccentricity. 
%
The crossing time of systems containing retrograde planets were studied by Smith and Lissauer (2009). 
They have found that 
systems with mixture of planets in retrograde and prograde orbits are more stable than the system which has only prograde planets provided that the total number of planets and their orbital separations in mutual Hill radii are the same. 

%
When planets embedded in protoplanetary gas disk, 
drag force which damps eccentricities also affects crossing time (Iwasaki $et\ al$., 2001, 2002; Iwasaki and Ohtsuki 2006). 
When the crossing time without drag force is shorter than eccentricity damping timescale, the drag force hardly changes the crossing time. 
Otherwise, the orbital crossing time is at least 200-times longer than that without the drag. 
This result implies that multiple planet systems do not start orbital instability until disk gas is sufficiently depleted. 
If the effects of type-I migration and disk inner edge are taken into account, the systems can become resonant. 
We will show that the resonant configuration stabilizes the systems even after disk gas is completely depleted, if the number of planets is smaller than a critical value.

\section{Numerical Model}\label{num_model}
We consider a situation that planets are brought to current locations near their host star by type-I migration, which leads the systems to resonant configuration in the protoplanetary disk. 
Planetary growth simulations including type-I migration and disk inner edge by Ogihara and Ida (2009) suggest that similar-sized 
planets are trapped in the resonances. 
We consider that planets have equal masses ($M_{\rm p}=3M_{\oplus}-30M_{\oplus}$) and coplanar orbits around the central star with $M_*=1M_{\odot}$ in all cases (non-zero inclination cases are studied in Yoshinaga $et\ al$. 1999). 
Using 4th-order Hermite scheme, we continue calculations until a distance between planets becomes smaller than their mutual Hill radius for at least one pair or until we reach to an upper limit of orbital evolution time. 
We set the upper limit of our calculation at $10^8$ Kepler time
 of the innermost planet ($a_1 = 0.1$ AU for all cases).
Planetary mass ($M_{\rm p}$), orbital separation normalized by mutual Hill radius ($K$), and the total number of bodies ($N$) are treated as parameters.

In this paper, we target on the first-order mean-motion resonances, i.e., planets have $p+1$:$p$ period relation, according to the results by Ogihara and Ida (2009). 
Planets which are in a mean-motion resonance have a relation between their pericenters and a point of conjunction. 
Even if whether planets have a periods ratio of $p+q$:$p$, it does not guarantee that they are in a mean-motion resonance. 
Thus, for the same Hill separations and planetary mass, we can set up both resonant and non-resonant systems. 
In the non-resonant cases, 
planets have the $p+1$:$p$ period ratio, but their initial longitudes are given randomly.
In the resonant cases, we put planets in the mean-motion resonance using orbital migration. 
The orbital migration automatically leads the planets to the mean-motion resonance with appropriate resonant angles. 
To extract the effect of a mean-motion resonance on the orbital stability, we compare the crossing time for the resonant cases with that for the non-resonant cases.

\subsection{Configuration of System}\label{conf_system}
Here, we explain how we control separations of neighboring planets in the same resonance using only one parameter $K$. 
Mutual Hill radius of the $i$-th planet and the $(i+1)$-th planet is given by
\begin{eqnarray}
 r_{{\rm H}i, i+1} = \left( \frac{M_i + M_{ i+1} }{3 M_*} \right)^{1/3} \left( \frac{M_i a_i + M_{ i+1} a_{ i+1} }{M_i + M_{ i+1}} \right), 
 \label{eq:mutual_Hill}
\end{eqnarray}
where $a_i$ is $i$-th planetary semi-major axis, $M_i$ is $i$-th planetary mass, and $M_*$ is stellar mass. 
Using a factor $K$, the orbital separation of neighboring planets is expressed as
\begin{eqnarray}
 a_{ i+1} - a_i = K r_{{\rm H}i, i+1}.
 \label{eq:equal_Hill}
\end{eqnarray}
In our simulations, all planets have an equal mass $M_i = M_{i+1} = M_{\rm p}$ and neighboring plantes have period ratio of $p+q$:$p$, i.e., $a_{i+1}/a_i = (p/p+q)^{-2/3}$. 
Using these relations, $K$ is expressed as 
\begin{eqnarray}
 K = \frac{2\{(p+q)^{2/3} - p^{2/3} \}}{(p+q)^{2/3} + p^{2/3}  }  \left( \frac{3 M_*}{2M_{\rm p}} \right)^{1/3}. 
 \label{eq:K_resonance}
\end{eqnarray}
Note that $K$ in the same for all the adjacent pairs for given $p$ and $q$. 
The values of $K$ we use are shown in Table \ref{table:cases}. 
Semi-major axis of the $i$-th planet ($i\geq2$) in the first-order mean-motion resonance is expressed as
\begin{eqnarray}
 a_{i} &=&  \left( \frac{ 2{\bar M} +  K  }  { 2{\bar M}- K } \right)^{i} a_1,
 \label{eq:ith_semi-major}
\end{eqnarray}
where ${\bar M} = \left( 2M_{\rm p} /3 M_* \right)^{-1/3} $. 
We use $a_1=0.1$ AU in all cases. 

\subsection{Non-Resonant Cases}
For non-resonant cases, we set planets according to Eq. \ref{eq:ith_semi-major} without any special treatment like resonant case (\S \ref{sec:r_case}). 
We integrate the equations of motion, 
\begin{eqnarray}
  \frac{d^2 \textrm{\boldmath $r$}_{ i}}{dt^2} = -{\rm G}M_{ *}\frac{\textrm{\boldmath $r$}_{ i}}{r_i^3} - \sum_{ j \neq i}^{ N} {\rm G}M_{\rm p} \frac{\textrm{\boldmath $r$}_{ ij} }{r_{ ij}^3} - \sum_{ j}^{ N} {\rm G}M_{\rm p}\frac{\textrm{\boldmath $r$}_{ j}}{r_{ j}^3}, 
  \label{eq:basic_commensurable}
\end{eqnarray}
where $i$ refers to the $i$-th protoplanet ($i=1, 2, \cdots, N$), G is the gravitational constant, and 
$\textrm{\boldmath $r$}_{ ij} $ is relative distance of the planet $i$ and $j$. 
We perform 100 runs for each value of $N$. 

\subsection{Resonant Cases}\label{sec:r_case}
We form exact resonance situations by orbital migration simulations in a gaseous disk. 
After all planets are locked in a resonance, we gradually deplete the gas. 
We calculate crossing times of 5 resonances. 
Choices of the resonance and planetary mass are in Table 1. 
By the choice, the Hill separation $K$ is automatically adjusted as we explained in \S \ref{conf_system}. 
In case1 and case2, we use typical values of resonant planets obtained in simulations by Terquem and Papaloizou (2007) and Ogihara and Ida (2009). 
The other sets are chosen to study how crossing time changes by these parameters. 
We calculate 10 runs in 6:5 and 7:6 mean-motion resonance, and 3 runs in other cases slightly changing initial longitude. 
We follow Ogihara and Ida (2009)'s settings.
The basic equations to form initial conditions are
\begin{eqnarray}
  \frac{d^2 \textrm{\boldmath $r$}_i}{dt^2} = -{\rm G}M_{*}\frac{\textrm{\boldmath $r$}_i}{r_i^3} - \sum_{ j \neq i} {\rm G} M_{\rm p} \frac{\textrm{\boldmath $r$}_{ ij} }{r_{ ij}^3} - \sum_j {\rm G} M_{\rm p} \frac{\textrm{\boldmath $r$}_j}{r_j^3}
 + \textrm{\boldmath $F$}_{\rm damp} + \textrm{\boldmath $F$}_{\rm mig}, 
 \label{eq:typeI}
\end{eqnarray}
where $ \textrm{\boldmath $F$}_{\rm damp}\ {\rm and}\ \textrm{\boldmath $F$}_{\rm mig}$ are the specific forces 
owing to eccentricity damping due to tides from the gas disk as a drag force (e.g., Tanaka and Ward 2004) and type-I migration, respectively. 
These force are given by Ogihara and Ida (2009) as 
\begin{eqnarray}
 \textrm{\boldmath $F$}_{\rm damp} &=& \left( \frac{M_p}{M_*} \right) \left( \frac{v_{\rm K}}{c_s} \right)^4 \left( \frac{\Sigma_g r^2}{M_*} \right) \Omega \left[
 \left\{ 0.114 (v_{\theta} - r\Omega ) + 0.176 v_r \right\} \textrm{\boldmath $e$}_{r} \right. \nonumber \\
 && \left.
 + \{ -1.736 (v_{\theta} - r\Omega) + 0.325 v_r\} \textrm{\boldmath $e$}_{\theta}
 +\{-1.088 v_z -0.871 z \Omega\} \textrm{\boldmath $e$}_{z}
 \right], \\
 \textrm{\boldmath $F$}_{\rm mig} &=& -2.17 f_{\rm m}\frac{M_p}{M_*} \left( \frac{v_{\rm K}}{c_s} \right)^2 \frac{\Sigma_g r^2}{M_*} \Omega v_{\rm K} \textrm{\boldmath $e$}_{\theta}, 
\end{eqnarray}
where $f_{\rm m}$ is a scale parameter corresponding to an uncertainty in type-I migration. 
%
%
These additional forces arise from interaction with the gas disk. 
For the disk model, we use 
\begin{eqnarray}
 \Sigma_{\rm g} &=& 2400 f_{\rm g} \left( \frac{r}{1{\rm AU}} \right)^{-3/2} {\rm g\ cm}^{-2}, \label{eq:Disk_SDensity}\\
 c_{\rm s} &=& 1.0 \times 10^5 \left( \frac{r}{1{\rm AU}} \right)^{-1/4} \left( \frac{L_*}{L_{\odot}} \right)^{1/8} {\rm cm\ s}^{-1},
\end{eqnarray}
where $\Sigma_{\rm g}$ is a surface density of the gas disk and $c_{\rm s}$ is the sound velocity.
During migration simulations, the surface density is 1.4 times larger than that of the Minimum Mass Solar Nebula model, i.e., $f_{\rm g}=1$. 
The sound velocity is that for an optically thin disk. 
We assume that the disk surface density smoothly vanishes with a hyperbolic tangent function at inner edge as 
\begin{eqnarray}
 \tanh{\left( \frac{r - r_{\rm edge}}{\Delta r} \right)},
\end{eqnarray}
where $r_{\rm edge}$ is a heliocentric distance of the inner edge of the gas disk, and $\Delta r$ represents typical width of the inner edge.
We choose $r_{\rm edge} =0.1$ AU and $\Delta r = 0.001$ AU. 
In our calculations, we put all planets slightly outside of $p+1$:$p$ resonance. 
Planets slowly migrate inward and are trapped at $p+1$:$p$ resonance automatically. 
After planets are captured in mean-motion resonances, we perform orbital integration for crossing time, decreasing gas density. 
For adiabatic gas depletion, we decrease $f_{\rm g}$ as
\begin{eqnarray}
 f_{\rm g} = \exp{\left(- \frac{t}{\tau_{\rm dep}} \right)},
 \label{eq:depletion}
\end{eqnarray}
where $\tau_{\rm dep}$ means depletion timescale. 
The timing $t=0$ is the starting time of gas depletion. 
For adiabatic gas depletion, we normally take $\tau_{\rm dep}=10^4$ yr following Ogihara and Ida (2009). 
We check the dependence of $\tau_{\rm dep}$ on the crossing time by changing it to $\tau_{\rm dep} = 10^4\ {\rm yr},\ 10^3\ {\rm yr},\ 10^2$ yr in \S \ref{sec_result_r}. 
The system is stable over $t_{\rm drag}$ (defined in Appendix \ref{sec:g_dep}) due to eccentricity damping. 
To distinguish the resonant effect from the stabilization effect due to eccentricity damping, 
we choose these timescales as $\tau_{\rm dep}$. 

The timescale of eccentricity damping obtained by linear calculation of tides from the gas disk (Tanaka and Ward 2004) is 
\begin{eqnarray}
 t_{\rm damp} 
 &\simeq& 0.96 f_{\rm g}^{-1} \left( \frac{M_{\rm p}}{10^{-5}M_{\odot}} \right)^{-1}  
 \left( \frac{a}{0.1{\rm AU}} \right)^{2} 
 \left( \frac{M_*}{M_{\odot} }\right)^{-1/2} \left( \frac{L_*}{L_{\odot}} \right)^{1/2} {\rm yr}, 
 \label{eq:e_damp}
\end{eqnarray}
where $M_{\odot}$ and $L_{\odot}$ are solar mass and luminosity. 
The timescale of semi-major damping by standard type-I migration (Tanaka $et\ al$. 2002) is
\begin{eqnarray}
 t_{\rm mig} 
 &\simeq& 4.8  \times 10^2 f_{\rm g}^{-1} f_{\rm m} \left( \frac{M_{\rm p}}{10^{-5}M_{\odot}} \right)^{-1}  \left( \frac{a}{0.1{\rm AU}} \right)^{3/2} 
 \left( \frac{M_*}{M_{\odot} }\right)^{1/2} \left( \frac{L_*}{L_{\odot}} \right)^{1/4} {\rm yr}. 
\end{eqnarray}
To stop the innermost planet at the disk edge of $a_1=0.1$ AU and to trap following planets in a resonances, 
we adopt slow enough migration, $f_{\rm m} \geq 50$ (Ogihara $et\ al$. 2010). 
When a planet is in the first-order resonance, the timescale of resonant libration is 
\begin{eqnarray}
 t_{\rm lib} 
 &=&
 11.7 
 \left(\frac{p}{5}\right)^{-1} \left(\frac{\alpha F_{\rm D} }{-3.94613}\right)^{-1/2} \left( \frac{e}{10^{-3}}  \right)^{-1/2} \left(\frac{a}{0.1{\rm AU}}\right)^{3/2} 
 \left( \frac{M_{\rm p}}{10^{-5}M_{\odot}} \right)^{-1/2} \nonumber \\ &&\times 
 E(\sqrt{\sin^2{\left(\varphi_0/2\right) }} ) \ {\rm yr}, \\
 \label{eq:lib_timescale}
 F_{\rm D} &=& \frac{1}{2} \left[ -2p-\alpha \frac{d}{d\alpha} \right] b_{1/2}^{(p)}, 
\end{eqnarray}
where $\alpha$ is the ratio of semi-major axis of the inner planet of the resonant pair divided by semi-major axis of the outer planet, $b_{s}^{(p)}$ is Laplace coefficient, 
$E$ is the complete elliptic integral of the first kind, and $\varphi_0$ is the maximum width of resonant libration (Murray and Dermott 1999). 
For adiabatic gas depletion, gas depletion timescale $\tau_{\rm dep}$ should be much longer than $t_{\rm lib}$. 
The gas depletion in this work is always adiabatic.

\section{Results}\label{results}

\subsection{Dependance of Crossing Time on Number of Planets}
\label{Num}
\subsubsection{typical orbital evolution}
In resonant cases, we make initial conditions by damping of $e$ and $a$, as explained in \S \ref{sec:r_case}. 
Fig. \ref{fig:semima_example} shows evolution of the semi-major axes of planets. 
The planets migrate inward due to type-I migration and are trapped in $7$:$6$ mean-motion resonances at $t\simeq-4500$ yr. 
Although planets are still subject to type-I migration, the innermost planet is caught in the disk edge at 0.1 AU and the other planets do not migrate furthermore. 
After gas depletion, planets start instability at $t\simeq 73000$ yr. 
We confirmed the mean-motion resonance from plain commensurability by resonant angle. 
When a planet is in $7$:$6$ mean-motion resonance, the resonant angle of outer planet is written as $\varphi^{\prime}_i = 7\lambda_{ i+1} - 6 \lambda_i - \varpi_{i+1}$, where $\lambda$ is the mean longitude and $\varpi$ is the longitude of pericenter and the resonant angle of inner planet is written as $\varphi_i = 7\lambda_{ i+1} - 6 \lambda_i - \varpi_{i}$. 
In the case of systems that are formed by migration (Fig. \ref{fig:phiout_dep_t1_76_5-4}), 
resonant angles of inner planet librate around $\varphi=0$ and resonant angles of outer planet librate around $\varphi^{\prime}=\pi$. 
Since $\varphi=0$ and $\varphi^{\prime}=\pi$, all conjunctions occur when inner planet is at its pericenter and outer planet is at its apocenter. 
This means that only one configuration is possible when $\varphi=0$ and $\varphi^{\prime}=\pi$. 
The resonant angles start circulation at about 70000 yr, just before the occurrence of instability. 
On the other hand, if we just put planets at semi-major axes of 7:6 resonance without such special treatment migration, resonant angles circulate and do not take a particular value (Fig. \ref{fig:phiout_5_5_7-6}). 
That is, this initial condition is non-resonant.

\subsubsection{non-resonant case}
In the non-resonant case, we put planets ($M_{\rm p}=10^{-5}M_{\odot}$) on the orbital separation of $\Delta a= a_{i+1} -a_{i} =6.450r_{{\rm H}i,i+1}\ (K=6.450)$. 
We calculate $N=3,\ 5,\ 8,\ 9,\ 10,\ 11,\ 20,\ 30,\ {\rm and}\ 50$ cases. 
Changing initial longitude randomly, we repeat the simulation 100 times for each $N$. 
Fig. \ref{fig:5-65_dep} shows crossing time of 6:5 case (case1) having the different number of planets. 
The crossing time is normalized by Kepler time of the innermost body at $a_1 = 0.1$ AU. 
The circles are the crossing time of non-resonant system. 
The solid curve is a least square exponential fit for the results of non-resonant system. 
Although there is some fluctuation, 
the crossing time decreases with $N$. 
However, in the region of $N\gtrsim10$, crossing time is almost constant, $t_{\rm cross} \sim10^4\ T_{\rm Kep}$. 
This tendency is consistent with the result of Chambers $et\ al$. (1996) using $10^{-7}M_{\odot}$ planets. 
We formulate the crossing time as a function of the total number of planets such as
\begin{eqnarray}
 \log{t_{\rm cross}} = g\exp{\left(-\log{N}\right)} +h = \frac{g}{N} +h, 
 \label{eq:number_dependence}
\end{eqnarray}
where $g$ and $h$ are constants. 
From a least square fit for the data in Fig. \ref{fig:5-65_dep}, $g=6.21$ and $ h=3.39$ (the solid curve of Fig. \ref{fig:5-65_dep}). 
We show in \S \ref{KandM} how these constants depend on the Hill separation and planetary mass.

\subsubsection{resonant case}\label{sec_result_r}

The results in resonant case (case1) are summarized in Fig. \ref{fig:5-65_dep}. 
Triangles, squares, and crosses are the crossing time with $\tau_{\rm dep} = 10^4\ {\rm yr}$, $10^3\ {\rm yr}$, and $10^2\ {\rm yr}$, respectively. 
Three dotted curves are gas stabilization timescale ($t_{\rm drag}$) which is defined in Eq. \ref{eq:t_gas} for $\tau_{\rm dep} = 10^4,\ 10^3,\ {\rm and}\ 10^2$ yr from the top. 
Symbols shown on the top horizontal axis indicate lower limits of crossing time, since crossing has not been detected within $10^8\ T_{\rm Kep}$. 
In Fig. \ref{fig:5-65_dep}, the crossing time of the resonant case is longer than that of the non-resonant case. 
Since the resonant cases are stabilized by resonant effect and eccentricity damping by gas, we have to estimate the timescale of the stabilization by gas to understand resonant effect. 
Gas drag is needed to form resonant systems, 
but its removal is required for the evaluation of the crossing time. 
Since rapid gas depletion makes system unstable, adiabatic gas depletion is needed. 
Due to the adiabatic depletion, planets are stabilized by gas on timescales $\sim t_{\rm drag}$ (defined in Appendix \ref{sec:g_dep}). 
In Fig. \ref{fig:5-65_dep}, $t_{\rm cross} \sim t_{\rm drag}$ for $N\gtrsim 10$. 
The gas drag effect is that $t_{\rm cross}$ cannot be shorter than $t_{\rm drag}$ suggested by Iwasaki $et\ al$. (2002). 
The crossing time in large $N$ is dependent on $\tau_{\rm dep}$. 
As $\tau_{\rm dep}$ becomes shorter, we can diminish the gas drag effect and $t_{\rm cross}$ approaches that in the non-resonant case for $N \gtrsim 10$. 
Then, we find that $t_{\rm cross}$ jumps up by several orders of magnitude at $N \sim 8$. 
This jump-up is due to resonant effect. 
In the following, we examine the critical total number of planets at which $t_{\rm cross}$ increase abruptly with decreasing $N$. 
Note that $N_{\rm crit}$ is almost independent of $\tau_{\rm dep}$. 
Since resonant libration timescale is about $20$ yr in case1, small $\tau_{\rm dep}$ is not long enough to guarantee adiabatic gas depletion in these cases. 
Therefore, we choose $\tau_{\rm dep} = 10^4$ yr in the following, although an off-set of $\sim t_{\rm drag}$ for $\tau_{\rm dep}=10^4$ yr is added in the results.

\subsection{Orbital Separation and Mass Dependence}
\label{KandM}
In this subsection, we show dependence of crossing time on the orbital separation in mutual Hill radii and the planetary mass. 
First, we change the orbital separation factor $K$ fixing planetary mass. 
It corresponds to a change of $p$ of the resonance. 
The results for 6:5 resonance ($K=6.450$), 7:6 resonance ($K=5.456$), and 8:7 ($K=4.727$) are shown in Fig. \ref{fig:5-65_dep}, \ref{fig:5-76}, and \ref{fig:5-87}, respectively. 
In non-resonant cases, $g$ and $h$ of Eq. \ref{eq:number_dependence} increase with increasing $K$. 
This tendency is consistent with Chambers $et\ al$. (1996). 
The crossing time of resonant planets shows discontinuity at a certain value of $N$ as explained in \S \ref{Num} for 6:5 resonance case. 
The critical number of holding planet is $N_{\rm crit}=4$ for 7:6 resonances and $N_{\rm crit}=3$ for 8:7 resonances. 
We find a tendency that $N_{\rm crit}$ decreases with increasing $p$ value of $p+1$:$p$ resonance, i.e., decreasing the orbital separation in mutual Hill radii, as we show it in Fig. \ref{fig:K_Ncrit}.

Next, we change planetary mass with fixed the orbital separation in mutual Hill radii. 
Since mutual Hill radius depends on planetary mass, we can choose some resonances having similar $K$ by changing planetary mass. 
We choose planets of mass $10^{-4}M_{\odot}$ for 4:3 resonances in case4.
In this case, the orbital separation in mutual Hill radii is equal to $K=4.715$ which is nearly equal to the case3 (8:7 resonance with $M_{\rm p} = 10^{-5}M_{\odot}$). 
These results are plotted in Fig. \ref{fig:4-43}. 
In non-resonant cases, crossing time is longer for larger masses. 
According to Chambers $et\ al$. (1996) and Zhou $et\ al$. (2007), the crossing time of non-resonant systems increases with increasing planetary mass for $K>4$ as long as planetary eccentricities are small enough. 
This tendency is also found in our resonant results. 
In resonant cases, $N_{\rm crit}$ is $7$ in case4. 
Since $N_{\rm crit}=3$ in case3, $N_{\rm crit}$ decreases with increasing $p$ value of $p+1$:$p$ or decreasing planetary mass. 

Quillen (2011) suggests that three-body resonance overlap affects crossing times of non-resonant systems. 
It would also be the case in our resonant systems. 
Our results show that $N_{\rm crit}$ increases with decreasing $p$ value. 
It is consistent with the fact that the resonance overlap less occurs with small $p$ value.

We find that a pair which causes instability is always the nearest neighbors in outer region for resonant cases. 
It is related to our exponential depletion of gas density expressed in Eq. \ref{eq:depletion}. 
Outer planets can cause instability while inner plants are still stabilized by gas. 
Planetary pair which causes instability depends on gas profile and the way of gas depletion.

\subsection{Chain of Resonance}\label{sec:chain}
We have been studying crossing time of planets which are in the same $p+1$:$p$ resonance for all the adjacent pairs. 
But there is not the cases for resonant planets that are observed and formed by N-body simulations of Ogihara and Ida (2009). 
For example, 
4 planets observed in KOI-730 have periods of 7.38469, 9.84978, 14.7845, and 19.72175 days, respectively (Lissauer $et\ al$. 2011). 
Lissauer $et\ al$. (2011) suggests that these planets are in a chain of resonance of 8:6:4:3. 
In this system, planets are in the first-order mean-motion resonant orbits with neighboring planets and with every other planet. 
To check orbital stability of such systems, we calculate the crossing time of systems whose planets are in $8$:$7$:$6$ chain resonance, repeatedly (case5). 
For example, in $N=4$ case, a 2nd innermost planet and a 3rd one are in $8$:$7$ resonance, and mean-motion ratios of planets are 28:24:21:18. 
When planets are in $8$:$7$:$6$ chain resonance, the pairs of every other planets are in 4:3 resonance. 
Fig. \ref{fig:5-876} shows that $N_{\rm crit}\sim6$. 
This is larger than $N_{\rm crit}=4$ of $7$:$6$ single resonance and $N_{\rm crit}=3$ of $8$:$7$ single resonance. 
Although chain resonance effect is unclear, there is possible that planets would be stabilized by 4:3 every other resonance ($N_{\rm crit}>7$ in $10^{-5}M_{\odot}$ planets) or that it would reduce crossing time dependence on $N$ 
like the case of mixture of planets in retrograde and prograde orbits (Smith and Lissauer 2009).

\section{Conclusions}\label{conclusions}

We have investigated the crossing time ($t_{\rm cross}$) of resonant systems by numerical simulations for 4:3, 6:5, 7:6 and 8:7 resonances. 
The crossing time of non-resonant plants decreases continuously with increasing the total number of planets $N$. 
In the case of resonant systems, however, while $t_{\rm cross}$ is comparable to that in non-resonant systems for large $N$, it abruptly changes for $N\leq N_{\rm crit}$. 
In that case, the resonant systems are stable during the simulation time ($10^8$ Kepler time). 
We examine 4 cases of different resonances with changing the orbital separation in mutual Hill radii  $K$ and planetary mass $M_{\rm p}$, and 1 case of chain resonances. 
When $K$ or $M_{\rm p}$ is fixed, 
$N_{\rm crit}$ increases as $p$ value of $p+1$:$p$ decreases. 
When planets of mass $10^{-5}M_{\odot}$ are in 6:5 resonances, $N_{\rm crit}=8$, when planets of mass $10^{-5}M_{\odot}$ are in 7:6 resonances, $N_{\rm crit}=4$, when planets of mass $10^{-5}M_{\odot}$ are in 8:7 resonances, $N_{\rm crit}=3$, and when planets of mass $10^{-4}M_{\odot}$ are in 4:3 resonances, $N_{\rm crit}=7$.
%
Observed planets are not always in a chain of the same resonance. 
We calculate 8:7:6 chain resonant case.
We find that 
$N_{\rm crit}=6$ of 8:7:6 resonance case is larger than either of 8:7 and 7:6 single cases.

In this paper, we set the innermost planet at 0.1AU, 
which is usually too far for tidal effect to be particularly important. 
However, many detected planets reside quite a bit closer to their host stars. 
As can be seen in the numerical simulations of Terquem and Papaloizou (2007), Papaloizou and  Terquem (2010), and Papaloizou (2011), 
tidal dissipation affects the stability of resonant systems due to two effects. 
One is eccentricity damping. 
Another is change of amplitudes of resonant libration. 
According to Papaloizou and Terquem (2010), the timescale of eccentricity damping due to the tidal dissipation is 
\begin{eqnarray}
 t^{\rm s}_{\rm e} 
 &=&1.818 \times 10^8  \left( \frac{M_{\oplus}}{M_{\rm p}} \right)^{2/3} 
 \left( \frac{20a}{\rm 1AU} \right)^{5} \left( \frac{Q^{\prime}}{50} \right) T_{\rm Kep}. 
\end{eqnarray}
The parameter $Q^{\prime}=3Q/2k_{\rm 2}$, where $Q$ is  the tidal dissipation function and $k_{\rm 2}$ is the Love number. 
Since planets are stabilized 
when $e$-damping timescale is shorter than the crossing time in gas free conditions, 
tidal dissipation stabilizes planets $a \lesssim 0.02$AU for our closely spaced systems. 
Tidal dissipation often increases amplitude of resonant libration in association with an increasing period ratio, $P_{i+1}/P_i$ (Terquem and Papaloizou 2007; Papaloizou 2011). 
Although tides help eccentricity damping in most cases, 
planets having large resonant amplitudes would easily move away from resonant configurations due to tides and cause instability. 

Many planets in observed multi-super Earths systems are orbiting near the central star. 
It is suggested that close-in super-Earths are formed through orbital migration of protoplanets and stopped near the disk inner edge due to resonant trapping (Terquem and Papaloizou 2007; Ogihara and Ida 2009). 
Even in 6:5 resonance, our simulation (\S \ref{KandM}) shows the system can hold 8 planets stably. 
If a planet is in $2$:$1$ mean-motion resonance, the orbital separation is larger than 10 Hill radius for a planet $M_{\rm p} <1.4 \times 10^{-4}M_{\odot}$. 
Observed systems composed of a few planets in 2:1 resonance are stable over $10^8$ Kepler time 
since these system can hold over 8 planets stably. 
The same goes for satellite systems such as Galilean satellites. 
On the other hand, observations find many systems that are not in a mean-motion resonance (Fabrycky $et\ al$. 2012). 
The systems of super-Earths far from the resonances could be formed by the scenario by Ogihara and Ida (2009) that planets more than the critical number are once trapped in resonance in gaseous stage and cause orbital instability after gas depletion.

\appendix
\section{Timescale of Gas Depletion}
\label{sec:g_dep}
%
%
When there is eccentricity damping force, a planetary system is stable (Iwasaki $et\ al.$ 2001, 2002), provided that $e$-damping timescale ($t_{\rm damp}$) is shorter than the crossing time in gas free case ($t_{\rm cross}^{*}$). 
Since we adopt the exponential decay of gas, $t_{\rm damp}$ exponentially increases and eventually exceeds $t_{\rm cross}^{*}$. 
We define $t_{\rm drag}$ as the timescale for $t_{\rm damp}$ to become longer than $t_{\rm cross}^{*}$. 
Using the formula by Iwasaki $et\ al$. (2002), we find 
\begin{eqnarray}
 t_{\rm drag} &\simeq& 2.30  b \left( K - K_{\rm drag}(t=0) \right)  \tau_{\rm dep},
 \label{eq:t_gas}
\end{eqnarray}
where
\begin{eqnarray}
 K_{\rm drag} &\simeq& \frac{1}{b} \log{\left( \frac{t_{\rm damp} }{T_{\rm Kep}} \right)} - \frac{c}{b}, 
 \label{eq:Iwasaki_crit}
\end{eqnarray}
$b$ and $c$ are defined in Eq. \ref{eq:Chambers+}.

\acknowledgments
We thank Masahiro Ogihara for useful comments. 
We also thank Konstantin Batygin and an anonymous referee for helpful comments to improve the paper.
This research was supported by a grant for the Global COE Program, ”From
the Earth to ”Earths””, MEXT, Japan.
M. N. was supported by MEXT-KAKENHI (21740324) Grant-in-Aid for Young Scientists (B).

\clearpage

\begin{table}[tp]
 \begin{center}
 \caption{Cases of our calculations.  \label{table:cases}}
 \begin{tabular}{ccccc}
  \tableline\tableline 
   & \ $p+1$:$p$\ resonance\   & \ planetary mass\  $M_{\rm p}$ & \ Hill separation\ $K$\ & \ critical number\ $N_{\rm crit}$ \\
  \tableline
  case1 & 6:5 & $10^{-5}M_{\odot}$ & 6.450 & 8\\
  case2 & 7:6 & $10^{-5}M_{\odot}$ & 5.456 & 4\\
  case3 & 8:7 & $10^{-5}M_{\odot}$ & 4.727 & 3\\
  case4 & 4:3 & $10^{-4}M_{\odot}$ & 4.715 & 7\\
  case5 & 7:6 and 8:7 & $10^{-5}M_{\odot}$ & 4.727 and 5.456 & 6\\
  \tableline
 \end{tabular}
\end{center}
\end{table}

\clearpage

\section*{Figure Captions}
\begin{figure}[tbhp]
  \begin{center}
   \includegraphics[width=170mm]{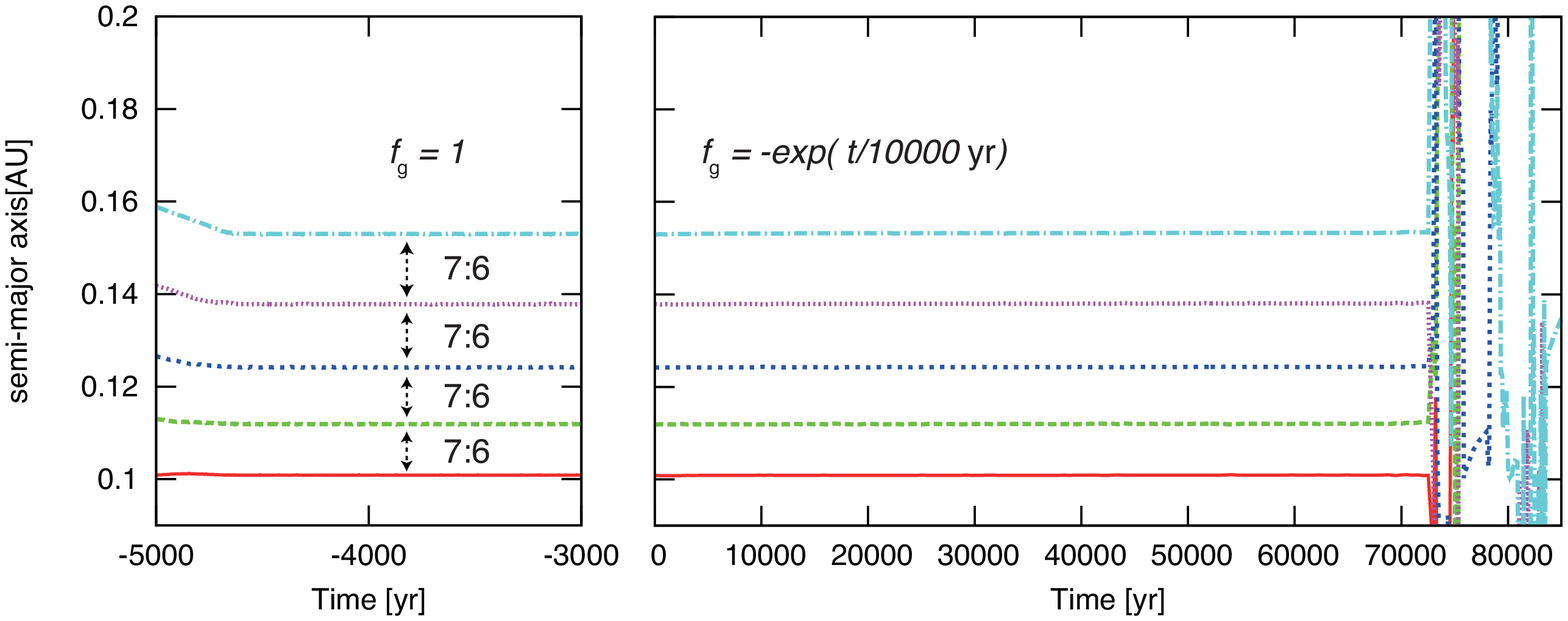}
  \end{center}
  \caption{}
  \label{fig:semima_example}
\end{figure}

Fig. \ref{fig:semima_example}. - An example of resonant trapping is shown. 
Time evolutions of semi-major axes. The total number $N$ is equal to 5. 
We form exact resonance situations by 5000 yr simulation of orbital migration 
before the reduction of the gas density at $t=0$. 
Planets migrate inward and are trapped at 7:6 resonance orbits. 
At $t=0$, we start gas depletion with depletion timescale $\tau_{\rm dep}=10^4$ yr. 
The crossing time is about 73000 yr ($\simeq 2.3\times 10^6\ T_{\rm Kep}$). 

\begin{figure}[htpb]
 \includegraphics[angle=-90,width=170mm]{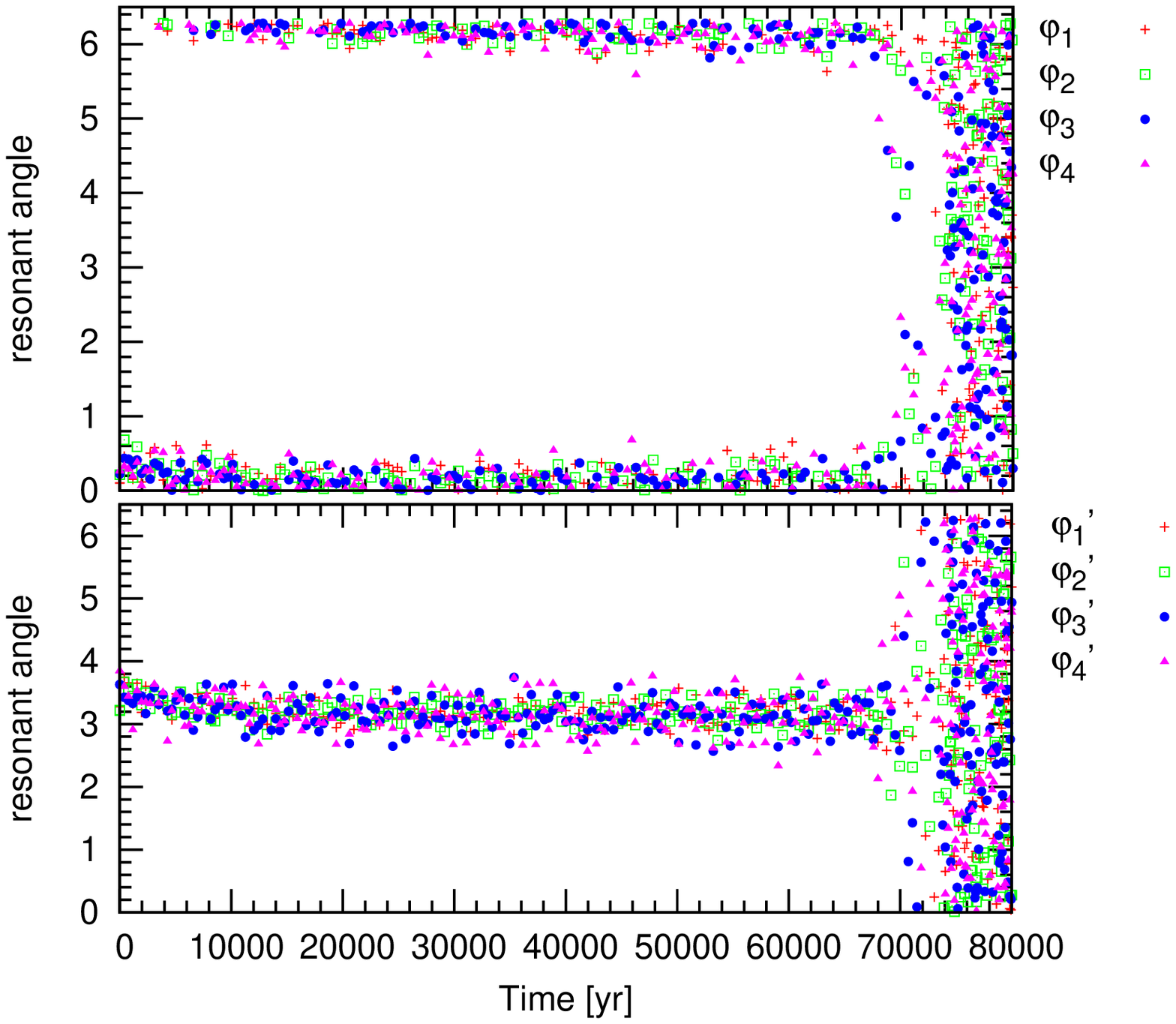}
 \caption{}
  \label{fig:phiout_dep_t1_76_5-4}
\end{figure}

Fig. \ref{fig:phiout_dep_t1_76_5-4}. - Time evolutions of resonant angle of 5 planets in 7:6 resonances. 
In this case, the orbital instability causes at about 73000 yr. 
Planets are labeled from the innermost.
The upper figure shows resonant angle $\varphi_1$ (cross), $\varphi_2$ (square), $\varphi_3$ (circle), and $\varphi_4$ (triangle). 
The lower figure shows resonant angle $\varphi_1^{\prime}$ (cross), $\varphi_2^{\prime}$ (square), $\varphi_3^{\prime}$ (circle), and $\varphi_4^{\prime}$ (triangle). 

\begin{figure}[htpb]
 \includegraphics[angle=-90,width=170mm]{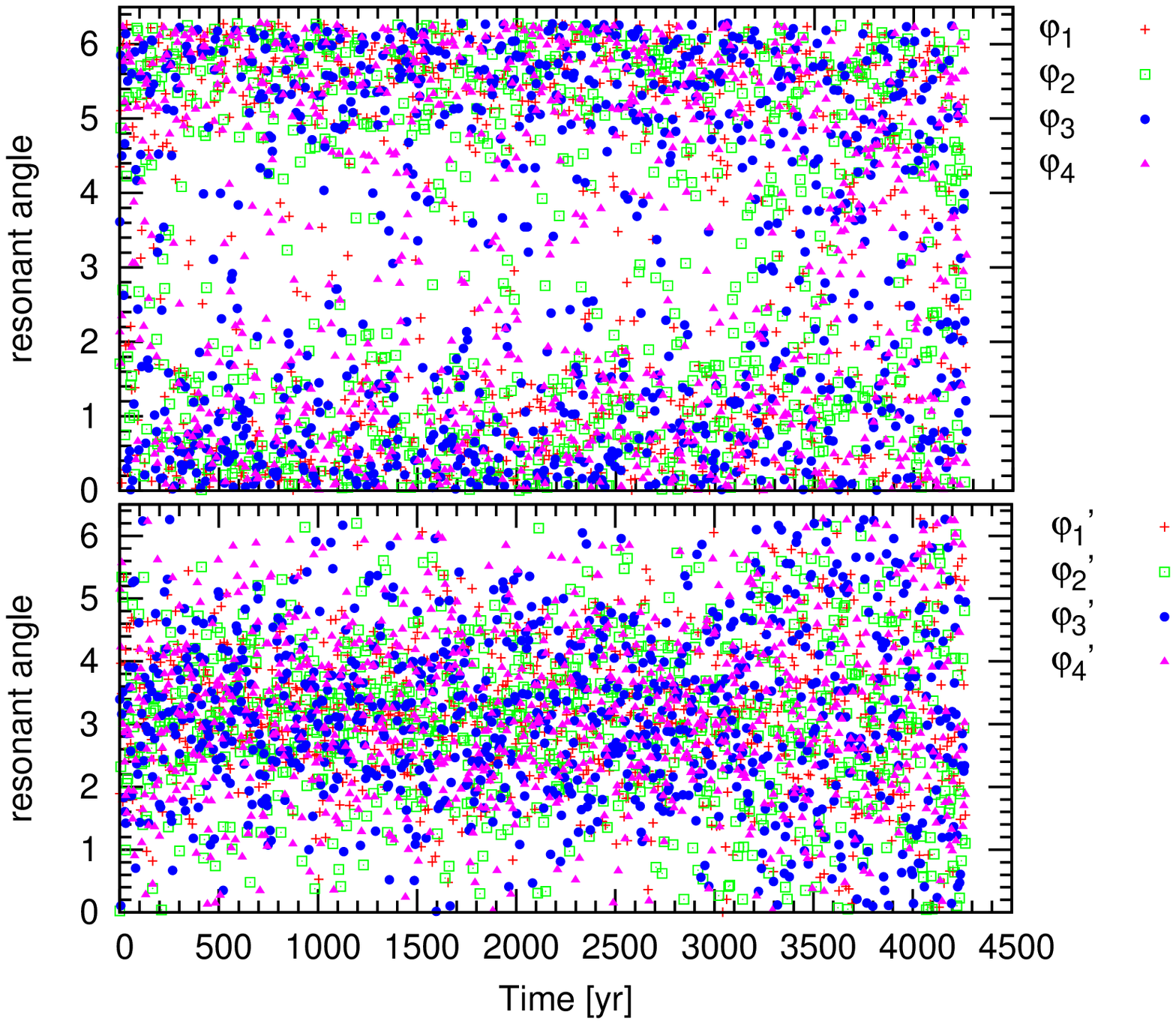}
 \caption{}
  \label{fig:phiout_5_5_7-6}
\end{figure}

Fig. \ref{fig:phiout_5_5_7-6}. - Time evolutions of resonant angles of non-resonant planets. 
This figure is the case of $N=5$ and $K=5.456$. 
The orbital instability causes at about 4200 yr. 
The upper figure shows resonant angle $\varphi_1$ (cross), $\varphi_2$ (square), $\varphi_3$ (circle), and $\varphi_4$ (triangle). 
The lower figure shows resonant angle $\varphi_1^{\prime}$ (cross), $\varphi_2^{\prime}$ (square), $\varphi_3^{\prime}$ (circle), and $\varphi_4^{\prime}$ (triangle). 

\begin{figure}[htpb]
\includegraphics[angle=-90,width=70mm]{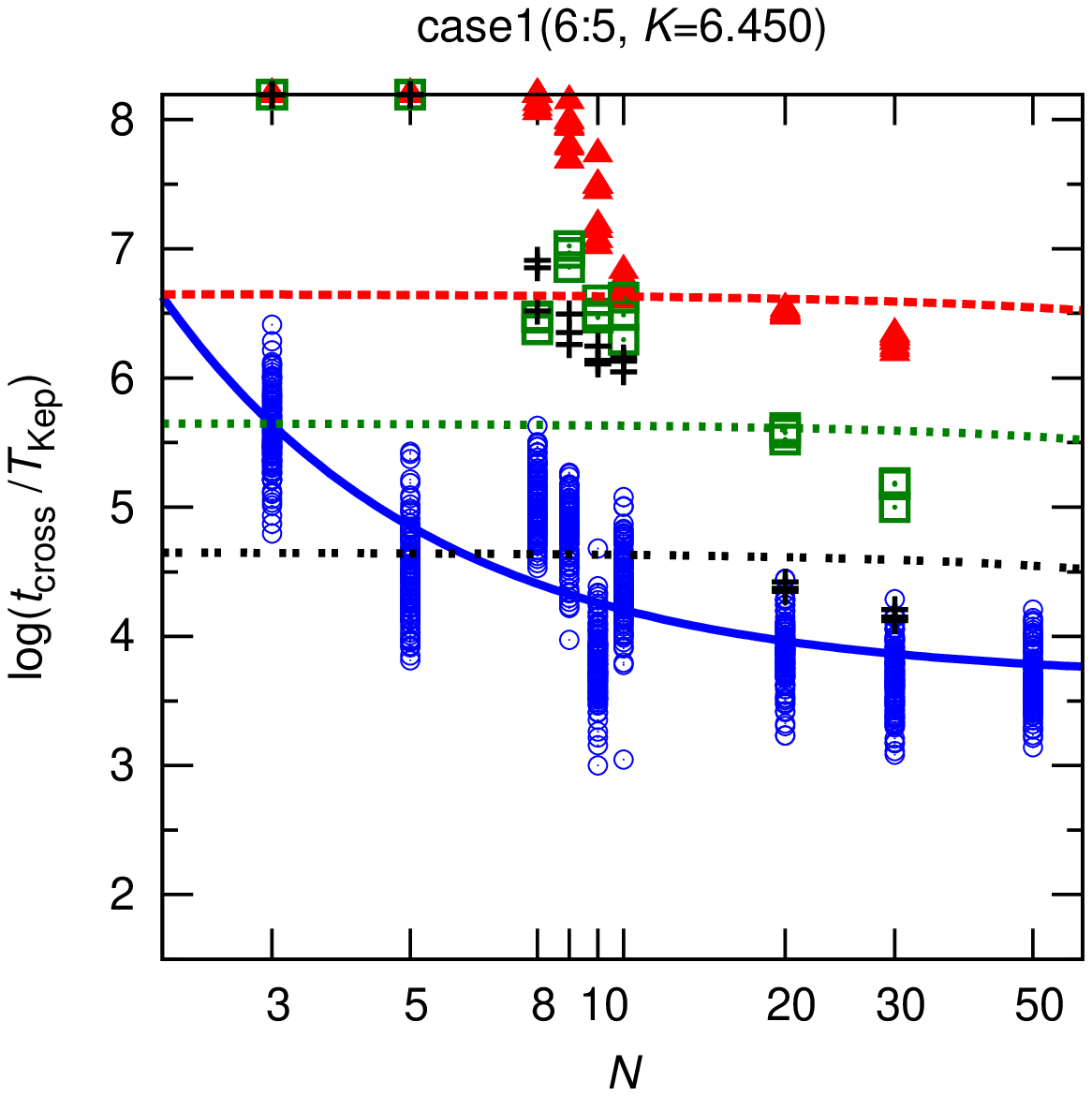}
\caption{}
\label{fig:5-65_dep}
\end{figure}

Fig. \ref{fig:5-65_dep}. - Crossing times $t_{\rm cross}$ normalized by Kepler time of the innermost body versus the number of planets $N$. 
The circles represent the crossing time of 6:5 orbits whose initial longitudes are chosen randomly (100 cases for each $N$). 
The solid curve is a least square exponential fit to the circles, $\log{t_{\rm cross}} = 6.21/N +3.39 $.
The triangles are the crossing time of 6:5 orbits (10 cases for each $N$) whose initial conditions are generated by orbital integrations including migration ($\tau_{\rm dep} = 10^4$ yr). 
The square symbols are $\tau_{\rm dep}=10^3$ yr and cross symbols are $\tau_{\rm dep}=10^2$ yr. 
Three dotted lines represent $t_{\rm drag}$ (Eq. \ref{eq:t_gas}) of the outermost planet for $\tau_{\rm dep} = 10^4,\ 10^3,\ {\rm and}\ 10^2$ yr from the top. 
In the case1, $N_{\rm crit}$ is equal to $8$.

\begin{figure}[htpb]
 \includegraphics[angle=-90,width=70mm]{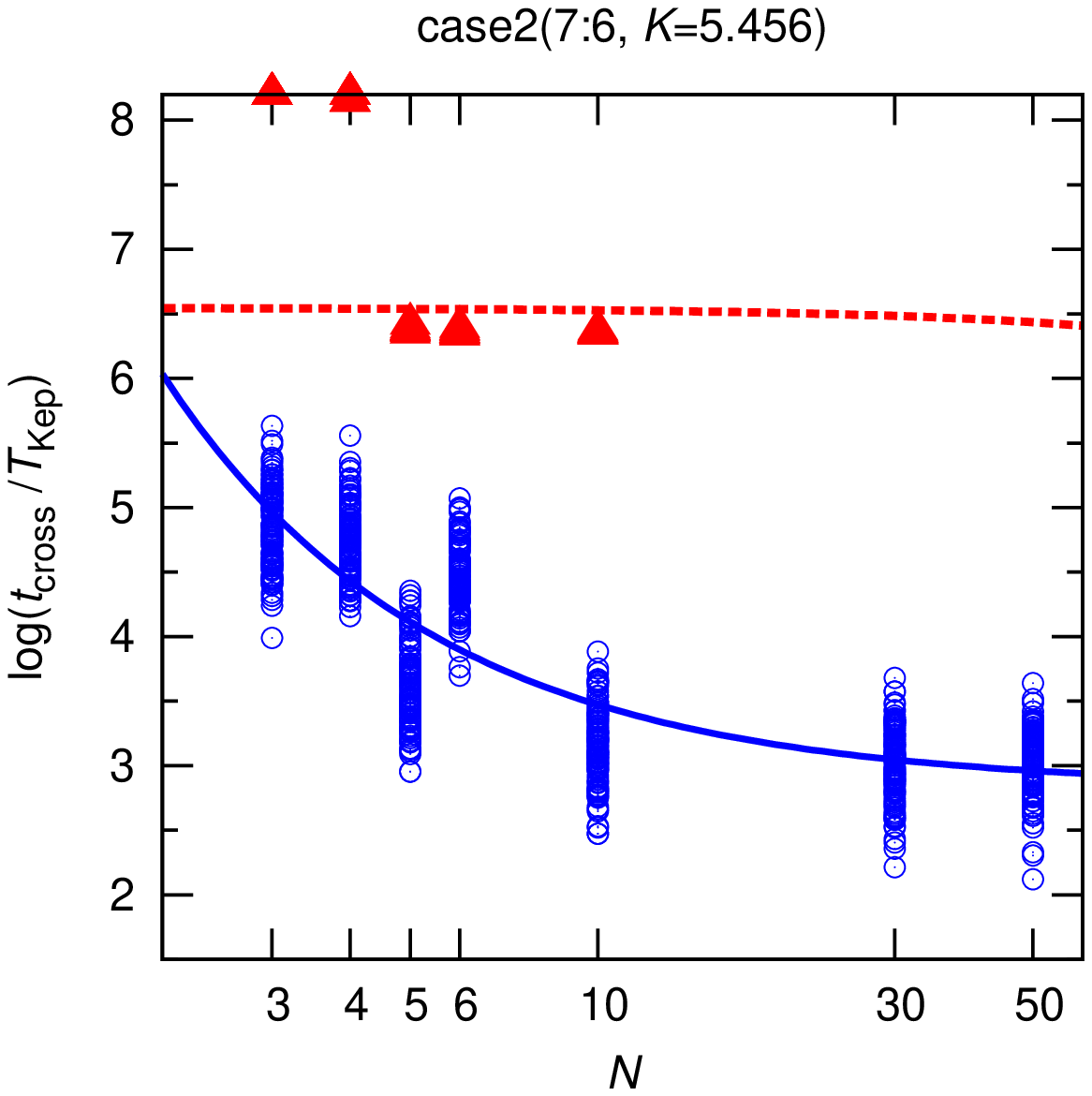}
 \caption{}
  \label{fig:5-76}
\end{figure}

Fig. \ref{fig:5-76}. - Same as Fig. \ref{fig:5-65_dep}, 7:6 resonances (case2). 
The circles represent the crossing time of 7:6 orbits whose initial longitudes are chosen randomly (100 cases for each $N$). 
The solid curve is a least square exponential fit to the circles, $\log{t_{\rm cross}} = 5.65/N +2.76 $.
The triangles are the crossing time of 7:6 orbits (10 cases for each $N$) whose initial conditions are generated by orbital integrations including migration ($\tau_{\rm dep} = 10^4$ yr). 
In the case2, $N_{\rm crit}$ is equal to $4$. 
The dotted line shows $t_{\rm drag}$ (Eq. \ref{eq:t_gas}) of the outermost planet. 

\begin{figure}[htpb]
\includegraphics[angle=-90,width=70mm]{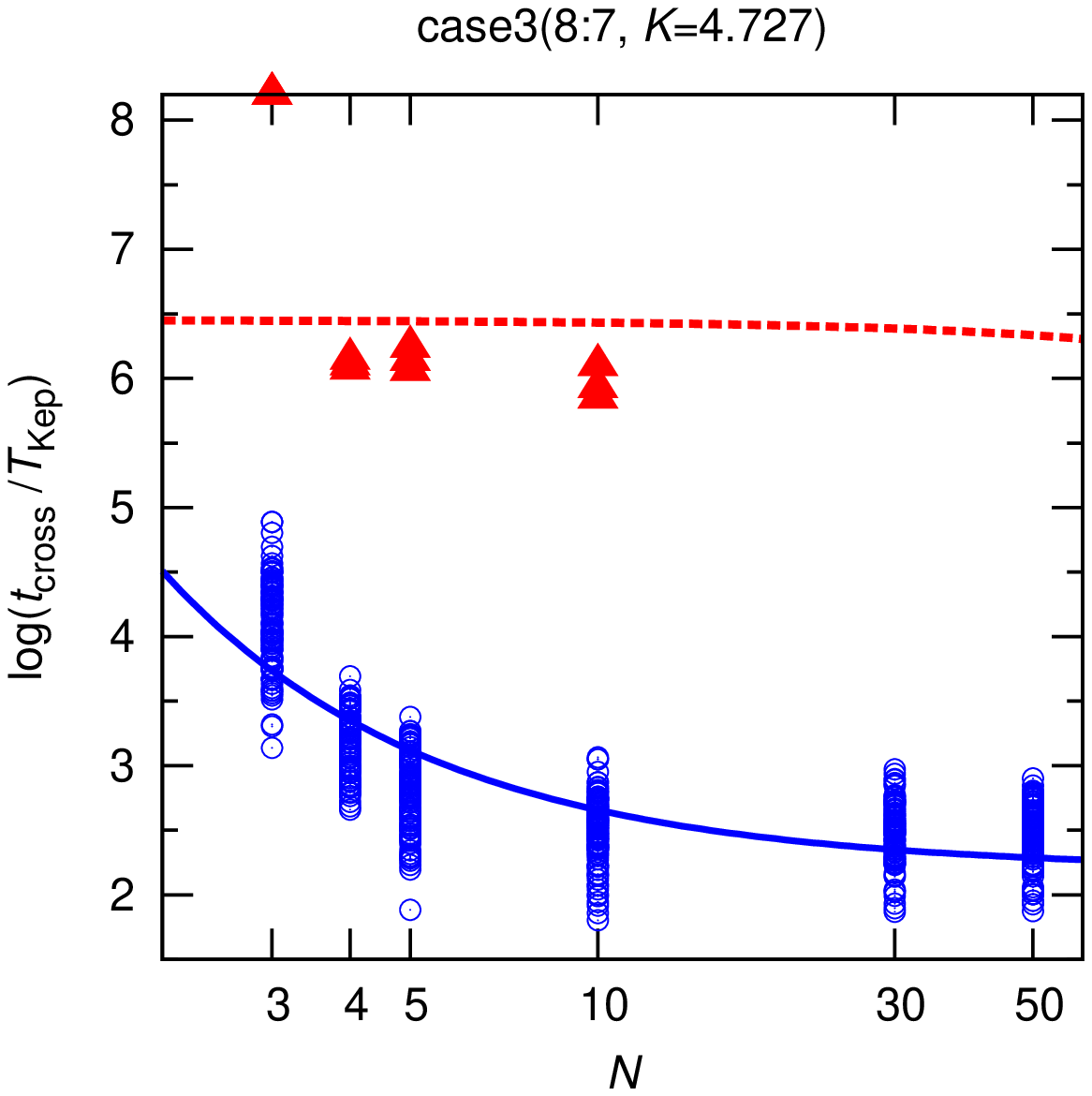}
\caption{}
\label{fig:5-87}
\end{figure}

Fig. \ref{fig:5-87}. - Same as Fig. \ref{fig:5-65_dep}, 8:7 resonances (case3). 
The circles represent the crossing time of non-resonant orbits. 
The triangles are the crossing time of resonant orbits (3 runs for each). 
The dotted line shows $t_{\rm drag}$ of the outermost planet. 
The solid curve is $\log{t_{\rm cross}} = 4.91/N +2.19 $.
Here, $N_{\rm crit}$ is equal to $3$.  

\begin{figure}[htpb]
\includegraphics[angle=-90,scale=.50]{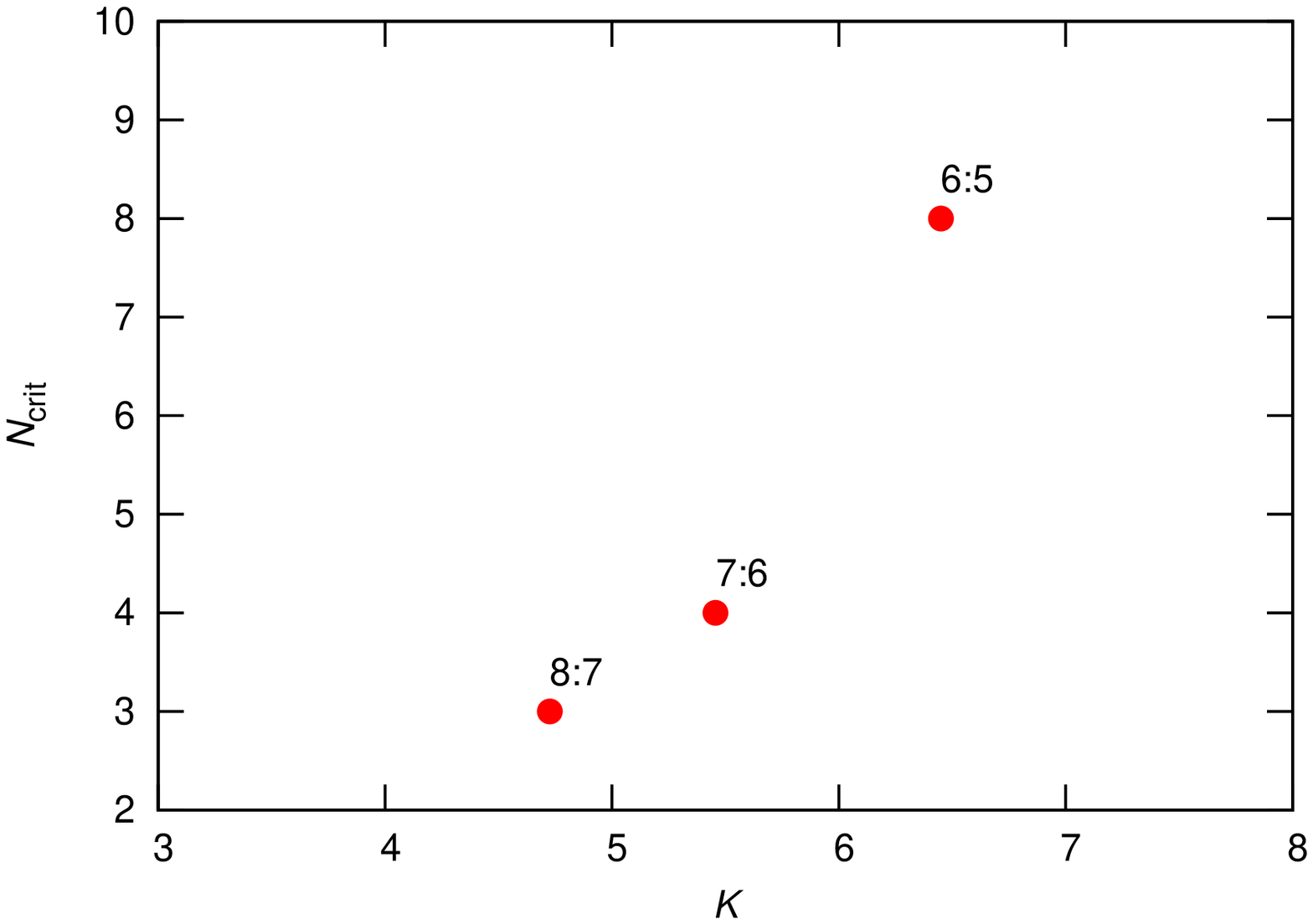}
\caption{}
\label{fig:K_Ncrit}
\end{figure}

Fig. \ref{fig:K_Ncrit}. - $N_{\rm crit}$ versus the orbital separation in mutual Hill radii $K$ in the case of $M_{\rm p} = 10^{-5}M_{\odot}$. 

\begin{figure}[htpb]
\includegraphics[angle=-90,width=70mm]{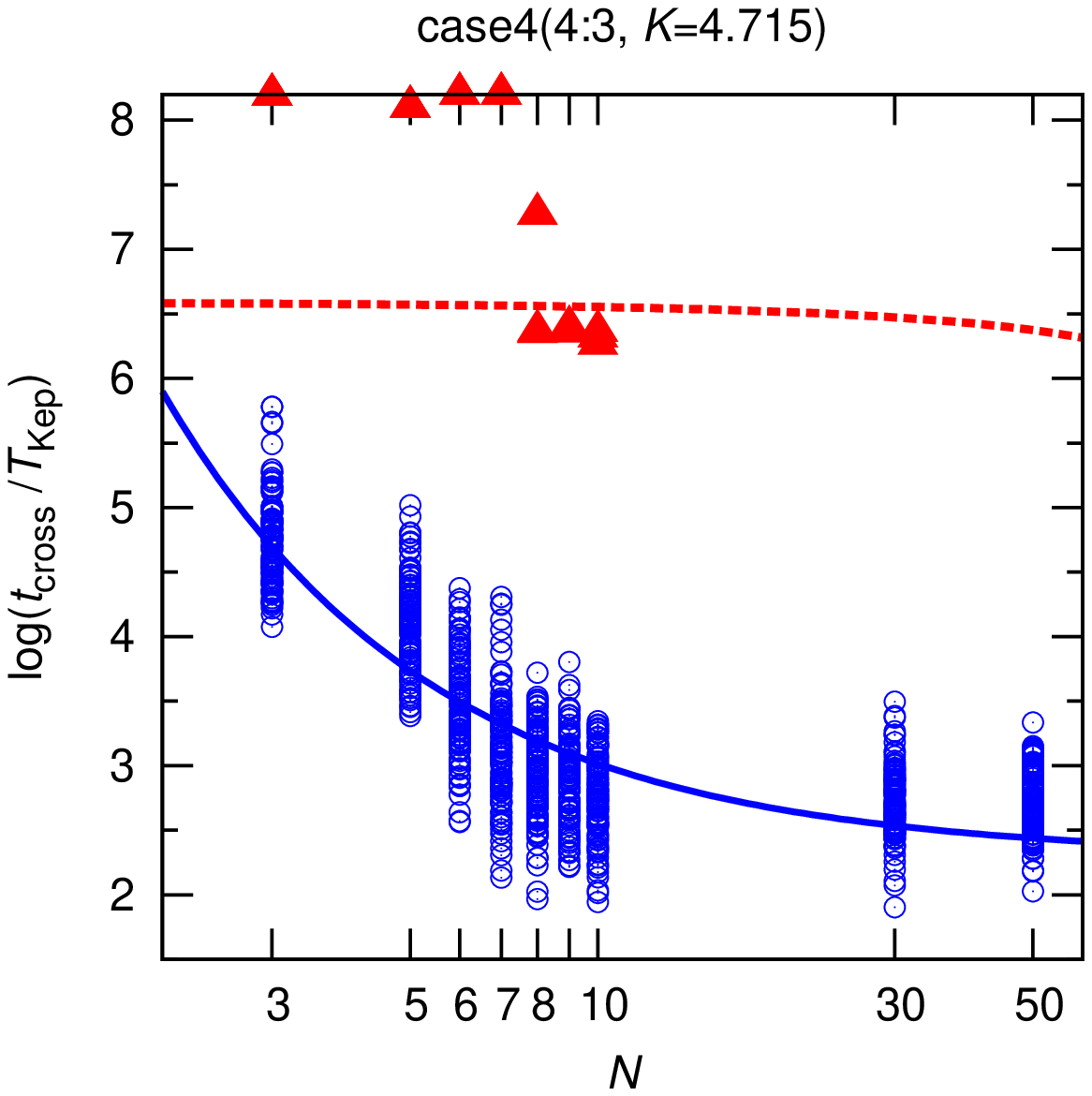}
\caption{}
\label{fig:4-43}
\end{figure}

Fig. \ref{fig:4-43}. - Same as Fig. \ref{fig:5-65_dep}, 4:3 resonances (case4).
The circles represent the crossing time of non-resonant orbits and the solid curve is a least square exponential fit to the circles, $\log{t_{\rm cross}} = 7.37/N +2.05 $.
The triangles are the crossing time of resonant orbits (3 runs for each). 
The dotted line shows $t_{\rm drag}$ of the outermost planet. 
In case4, $N_{\rm crit}$ is equal to $7$. 

\begin{figure}[htpb]
\includegraphics[angle=-90,scale=.50]{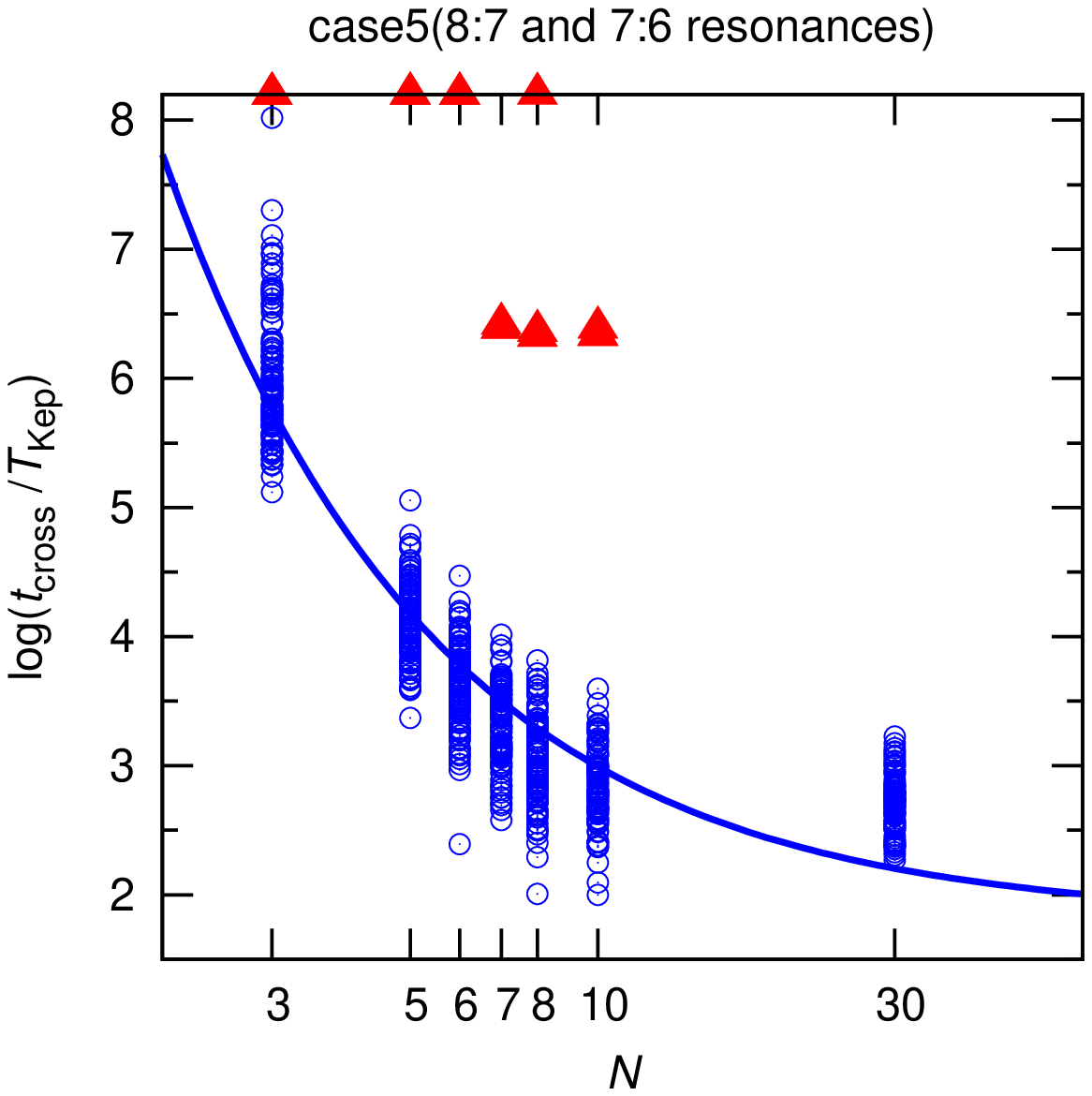}
\caption{}
\label{fig:5-876}
\end{figure}

Fig. \ref{fig:5-876}. - Same as Fig. \ref{fig:5-65_dep}, but for 8:7:6 resonances (case5).
The circles represent the crossing time of non-resonant orbits and the solid curve is a least square exponential fit to the circles, $\log{t_{\rm cross}} = 11.85/N + 1.81$.
The triangles are the crossing time of resonant chain orbits (3 runs for each). 
In case5, $N_{\rm crit}$ is equal to $6$.


\clearpage

\end{document}